\documentclass{svmult}

\usepackage{makeidx}         
\usepackage{graphicx}        
\usepackage{multicol}        
\usepackage[bottom]{footmisc}

\usepackage{url}
\usepackage{multirow} 

\makeindex             

\begin{document}
\hyphenation {O-me-mo-Con-di-tio-nal-Viz-For
              e-qui-va-lent-Class
              e-qui-va-lent-Property
             }

\title*{VPOET: Using a Distributed Collaborative Platform for Semantic Web Applications}
\titlerunning{Using a Distributed Collaborative Platform for Semantic Web Apps.}
\author{
Mariano Rico\inst{1}\and
David Camacho\inst{1} \and
Oscar Corcho \inst{2}
}
\institute{Escuela Polit\'{e}cnica Superior. UAM. \texttt{\{Mariano.Rico, David.Camacho\}@uam.es} \and 
Ontology Engineering Group. Departamento de Inteligencia Artificial. UPM. \texttt{ocorcho@fi.upm.es}}
%
%
\maketitle

This paper describes a distributed collaborative
wiki-based platform that has been designed to facilitate the development of
Semantic Web applications. The applications designed using this platform are able to
build semantic data through the cooperation of different developers and to exploit that semantic data.
The paper shows a practical case study on the application VPOET\index{VPOET}, and how an  application based on Google Gadgets has been designed to
test VPOET and let human users exploit the semantic data created. This
practical example can be used to show how different Semantic Web
technologies can be integrated into a particular Web application, and how the
knowledge can be cooperatively improved. 


\section{Introduction}
\label{Introduction}

One of the key aspects of the Semantic Web~\cite{bernerslee2001sw,Herman2006} is that software agents or applications are able to \textquotedblleft
understand\textquotedblright  the meaning of contents specifically designed for
them. The Semantic Web is made possible using a set of standards like RDF(S)~\cite{specRDF, specRDFS}\index{RDF}\index{RDFS}, OWL~\cite{Bechhofer2004}\index{OWL}, or SPARQL~\cite{specSPARQL}\index{SPARQL}, among others. 

In the Semantic Web research area, the concept of \textit{semantic information}
represents knowledge that can be automatically analysed with no (or minimal)
ambiguity. To avoid any possible ambiguity, the Semantic Web standards have been
designed using logic-based formalisms and ontological representations. For
example, there are a set of Description Logic \textit{reasoners} that can be used to perform inferences with OWL models. On the other hand, different knowledge standard representations,
named ontologies\index{ontologies}, have been designed to formally describe the exact meaning of a particular concept. An \textit{ontology} is a set of formal definitions about a particular domain. Although there exist other standards and formalisms to represent ontologies, the most popular in the Web is OWL which is based in the definition of \texttt{classes}, \texttt{properties}, \texttt{individuals}, and \texttt{relationships} between them. For example, the Friend Of A Friend(FOAF)\index{FOAF} ontology  can be used to define the Person and Organization classes; the name, surname and email properties; and the \textit{knows} relationship (applicable to individuals belonging to the Person class). The FOAF ontology comprises definitions, that is, no instances are declared for any defined class. Ontologies and data are identified by a namespace.

The evolution of the Semantic Web is directly joined to ontologies
and semantic technologies success. There are currently about 11,000
ontologies available on the Internet~\cite{warren2005, ding2004}, and the
semantic data has experimented an exponential growth for the last ten
years~\cite{Finin2006xtech}. However this high-quality information remains
hidden to most end-users, developers, and even software agents, because there
are only some few applications able to manage with this semantic data. Two main
problems can be analysed to explain this current situation. On the one hand, the
increasing difficulty to design adaptable and easily reusable Web applications
where a wide set of Web technologies and programming languages, such as HTML\index{HTML}, Javascript\index{Javascript}, CSS\index{CSS}, DHTML\index{DHTML}, Flash\index{Flash}, or AJAX\index{AJAX}, need to be
used, converting graphical-designers in skilled programmers as pointed in ~\cite{Rochen2006EUD}. On the other hand, the complexity of Semantic Web
technologies requires a very specialised knowledge. For instance, the
process of creating ontologies using OWL needs from domain experts and OWL
specialists in order to \textquotedblleft transfer\textquotedblright  the
experts' know-how into a specific OWL ontology. Therefore, the correct design of
a semantic web application needs from a wide set of different specialised experts.

This paper proposes a new approach to solve some of the previous problems. Our
approach is based on a particular methodology used to simplify the creation of Semantic Web Applications using a wiki-based approach, one of the most successful collaborative environments for the last years. Unlike common wikis, oriented to contents creation, some wikis can be used to functionality creation, in a collaborative way for developers.

This paper is structured as follows. Section~\ref{sec:metodology} shows our
methodological approach to design semantic applications based on wiki
technologies. Section~\ref{sec:using_vpoet} describes VPOET, a semantic application that implements the previous methodology. Section~\ref{sec:using_vpoet_http_channel} describes a
practical case study that exploits the communications channel provided by VPOET. Section~\ref{sec:matching} shows how to get the best fitted visualisation of a semantic data element for a given user profile. Finally, Section~\ref{sec:conclusions_futWork} summarises the conclusions and future work.

\section{Distributed Methodology for Semantic Cooperative-based Web applications}
\label{sec:metodology}
Interaction with human users, showing semantic data, or requesting data that will have to be converted to semantic data, is a cornerstone of the Semantic Web. Our work focuses on a technological approach, providing developers with a simple and collaborative programming framework in order to simplify the process of creation of semantic web applications. As a proof-of-concept, we present a real semantic web application that uses the aforementioned framework in order to validate the technological approach. Next subsections give the detail of this approach and a concrete implementation. 

\subsection{Designing a platform based in contribution for semantic applications developers}
Unlike recent efforts to create wiki-based technologies that allow editing semantic data (so-called semantic wikis, like Semantic Mediawiki, IkeWiki, or ODEWiki) in our approach we go a little bit further and allow users to create easily and collaboratively pieces of code that can be included in Semantic Web applications. This technological approach does not require developers with skills in multiple languages and technologies, but just wiki essentials, and basic skills on a programming language and semantic web technologies. For this kind of developers, and for a concrete wiki-engine called JSPWiki\footnote{See \url{http://jspwiki.org}}\index{JSPWiki}, we have created a software framework called Fortunata \index{Fortunata}. This software exploits plugins, software pieces that extend a given functionality. In this case, our plugins extend the functionality of an open-software wiki.
 Applications designed under this architectural paradigm let developers to create functionality in a decentralised way. Traditional development centralises the source code. Therefore, extending functionality typically requires accessing the source code and compile. The result is a new version of the application. However, plugins let members of a community to contribute creating new functionality with a minimal degree of dependence. When a developer has created and tested a new plugin, the source code is sent to the wiki administrator. If the code is considered valid and safe, it is compiled and added to the wiki engine. Unlike traditional development environments, this addition does not require to check for dependencies or compiling the whole application code. Even, in our system, it can be done while the application is running. 
Semantic web technologies provide us an additional advantage: simpler data integration. Fortunata-based applications comprise a set of plugins managing a semantic data source. These applications can integrate easily semantic data from other Fortunata-based applications.

\subsection{Applying the architectural aspects to real applications}
\begin{figure*}[t]
    \centering
    \includegraphics [width=4in]{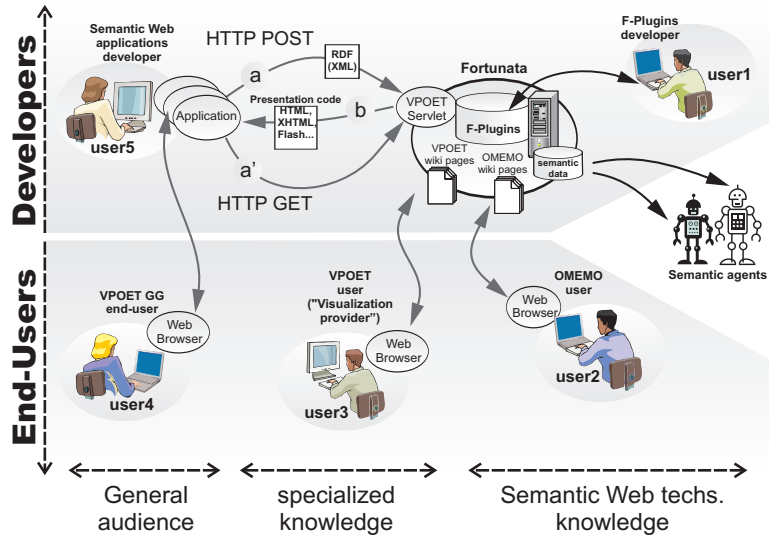}
    \caption{Involved roles in the proposed system}
    \label{fig:fortunata_parties}
\end{figure*}

As a result of applying this aspect, different roles appear for both developers and end-users. Figure \ref{fig:fortunata_parties} shows a clear separation between end-users, developers, and semantic agents, as well as different roles that are introduced below. 

The architectural aspect results in two different kinds of developers, as are shown in figure \ref{fig:fortunata_parties}. Table \ref{tabla:fortunata_roles} shows the activities and requirements of these users. User1 plays the role of \textquotedblleft semantic web applications developer\textquotedblright , providing with Fortunata-based plugins (F-plugins in figure \ref{fig:fortunata_parties}). A different kind of developer is represented by user5. She does not contribute with plugins, but takes advantage of the semantic data created by user1's applications.

As a proof-of-concept, we have created some Fortunata-based applications. In this paper we focus on VPOET. Let us see a brief description of this application and how it benefits from the methodological aspect.

VPOET enable end-users, denoted \textquotedblleft visualisation providers\textquotedblright  in this context, to create visualisation templates for a given ontology element, not only to show semantic data (output templates) but to request data from the user (input templates). These templates can be created by any user with basic skills in client-side technologies, such as HTML or Javascript, using simple macros provided in VPOET. Visualization providers can get information about the ontology element reading the wiki pages generated by another Fortunata-based application, or reading other manually created wiki pages referencing to these pages. In figure \ref{fig:fortunata_parties}, user3 represents this kind of user.

Besides creating the visualisation template, visualisation providers indicate the features of their templates using forms, specifying details such as template type (input or output), behaviour in case of changes to the font size, sizes (preferred, minimum, maximum), code-type provided (HTML, Javascript, CSS), or dominant colours. 
As any other Fortunata-based application, all the generated information is published as semantic data, so that it can be used by semantic agents. Besides, a HTTP GET/POST channel has been created to get access to the semantic data. Figure \ref{fig:fortunata_parties} shows this channel in the case of VPOET, and how it is exploited by developers like user5. For testing purposes, we have exploited this channel creating a Google Gadget\index{Google Gadget} called GG-VPOET. End users like user4 use GG-VPOET to render a semantic data source under a concrete visualisation template. 
Other applications can exploit this channel. For example, we are using this channel to query for the most appropriated visualisation for a given user profile. This experimental user profile contains data about the interactive impairments of the user, its interaction device, or its aesthetic preferences.

\begin{table}[!t]
    \caption{Description on the roles in the proposed system}
    \label{tabla:fortunata_roles}
    \centering
    \begin{tabular}{p{1cm} p{6cm} p{0.3cm} p{4cm}}
        \hline\noalign{\smallskip}
        \textbf{Role} & \textbf{Activities} & & \textbf{Requirements}\\
        \noalign{\smallskip}\hline\noalign{\smallskip}
        \raisebox{-1.5ex}[0pt]{user1} & F-plugins developer. Uses the Fortunata framework to create semantic plugins & & Basic java programming skills\\
        \noalign{\smallskip}\hline\noalign{\smallskip}
        \raisebox{-1.5ex}[0pt]{user5} & Semantic Web applications developer. Uses the HTTP channel provided by VPOET & & Basics of HTTP in any programming language\\
        \noalign{\smallskip}\hline\noalign{\smallskip}
        \raisebox{-3ex}[0pt]{user2} & OMEMO user. Any user interested in obtaining a simple and textual description of the elements in a given ontology & & \raisebox{-3ex}[0pt]{None}\\
        \noalign{\smallskip}\hline\noalign{\smallskip}
        \raisebox{-1.5ex}[0pt]{user3} & VPOET user. Client side graphical designer & & Requires basics of client side technologies\\
        \noalign{\smallskip}\hline\noalign{\smallskip}
        \raisebox{-3ex}[0pt]{user4} & VPOET-GG end-user. Any user interested in providing a visualisation of a semantic data source & & \raisebox{-3ex}[0pt]{None}\\
        \noalign{\smallskip}\hline
    \end{tabular}
\end{table}


\section{Using VPOET}
\label{sec:using_vpoet}
VPOET lets users create visualisation templates for any ontology element. Although VPOET can be used by any user with basic skills in client side-side web technologies, it has been created to let professional \textbf{graphical-designers} author attractive designs capable of rendering semantic data. Users of VPOET are denoted \textquotedblleft visualisation providers\textquotedblright  (VPs).
From an end-user point of view, this application is like any other web application, with form elements like text fields, radio buttons, or buttons. VPs just have to follow an online tutorial to start creating templates.

The process to create a template starts targeting an ontology element. For example, the next subsection reports on a use case that follows the tutorial aforementioned, in which the element \texttt{Person} from the FOAF ontology version 20050403 is targeted. The process to create the template comprises these steps:
\begin{enumerate}
  \item Getting information about the structure of the targeted element. That is, to know which sub-elements comprise the element. The visualisation provider obtains this information reading wiki pages automatically generated by OMEMO\index{OMEMO} (user2 in figure \ref{fig:fortunata_parties}), other Fortunata-based application.
   \item Authoring a graphical design in which semantic data will be inserted. End-users are free to use their favourite web authoring tool.
   \item Choose an identifier (ID) to create a wiki page with that ID. This wiki page shows information about the VP and its templates stored.
   \item The graphical design comprises a set of files (images, and client-side code such as HTML, CSS, or javascript). The client-side code is copied-pasted in the appropriated form fields. Image files or\textquotedblleft included\textquotedblright  files must be uploaded to the provider wiki page, or uploaded to any web server. In any case, the client code must point correctly to these files.
   \item A test loop starts, using semantic-data sources (typically external to VPOET) containing instances of the targeted element.
      \begin{enumerate}
          \item Paths (relatives or absolutes) must be substituted by means of a specific macro.
          \item Semantic data are inserted using specific macros.
          \item The design is tested against the test data sources
          \item This loop finish when the design produces a successful visualisation for all the semantic test data sources.
      \end{enumerate}
   \item The design is characterized by its creator, providing info about the template features, such as type, colors, size policy, or font changes behavior.
\end{enumerate}

Most of the effort required to create a template is located in the test loop, especially in the insertion of macros. The table \ref{tabla:vpoet_macros} shows the most relevant macros available in VPOET, the arguments each macro requires, and a brief explanation of each macro. 

%
%
%
%

\begin{table*}[!t]
    \caption{Main macros available for visualisation providers in VPOET.}
    \label{tabla:vpoet_macros}
    \centering
    \begin{tabular}{l p{2.5cm} p{6cm}}
        \hline\noalign{\smallskip}
        \textbf{Macro} & \textbf{Arguments} & \textbf{Explanation}\\
        \noalign{\smallskip}\hline\noalign{\smallskip}
        \texttt{OmemoGetP} & \texttt{propName} & It is substituted by the property value \texttt{propName}\\
        \hline
        \texttt{OmemoBaseURL} & No arguments & It is substituted by the URL of the server in which VPOET is running\\
        \hline
        \texttt{OmemoConditionalVizFor} & \texttt{propName, designerID, designID} & Renders the property \texttt{propName} only if it has a value, using the template indicated\\      
        \hline
        \texttt{OmemoGetLink} & \texttt{relationName} & It is substituted by a link capable of displaying elements of the type pointed by the relation \texttt{relationName}\\
        \noalign{\smallskip}\hline
    \end{tabular}
\end{table*}

VPOET has been designed to let its users reuse their templates. This is achieved using: (1) the conditional rendering of a property (using the macro \texttt{OmemoConditionalVizFor}) and (2) links capable of displaying the destination element of a relation (macro \texttt{OmemoGetLink}). A detailed explanation, and usage examples, can be found at \url{http://ishtar.ii.uam.es/fortunata}.

\section{Using the HTTP channel in VPOET}
\label{sec:using_vpoet_http_channel}
Although the information stored in VPOET is published as semantic data reachable through an URL that can be used by semantic agents, an additional channel to let non-semantic users access this information has been created. It has been implemented as a servlet that let users make HTTP GET/POST requests with variable parameters in order to facilitate queries like \textquotedblleft get an output visualisation created by provider X for the element foaf.Person.20050603 for the semantic data at URL Y\textquotedblright . The complete syntax is shown in Table \ref{tabla:vpoet_http}.

\begin{table*}[!t]
    \caption{Parameters accepted in the HTTP GET/POST request.}
    \label{tabla:vpoet_http}
    \centering
    \begin{tabular}{|p{2cm}|p{2.7cm}|p{7cm}|}
        \hline
        \textbf{Parameter} & \textbf{Value} & \textbf{Explanation/Example}\\
        \hline \hline
        \multirow {2}{*} \texttt{action} & \texttt{renderOutput} & Request a visualisation for the elements \texttt{object} in the data source given in parameter \texttt{object} \\
                                         & \texttt{renderInput}  & Request a visualisation to request data for the element \texttt{object} from the user\\
        \hline
        \multirow {2}{*} \texttt{object} & prefix.class[.ver]     & Example: foaf.Person \\
                                         & prefix.relation[.ver]  & Example: foaf.firstName \\
        \hline
                         \texttt{source (GET only)} & URL & URL of the semantic data source\\
        \hline
                         [\texttt{provider}] & ID & Identifier of the visualization provider. For example: user3.test\\
        \hline
        \multirow {3}{*} \texttt{outputFormat}  & \texttt{HTML} & Default value \\
                                                & \texttt{XHTML} & XHTML is used by WAP 2.0 mobile phones\\
        \hline
                         [\texttt{userProfile}] (GET only) & URL & URL of the RDF data source with the user profile\\
        \hline
    \end{tabular}
\end{table*}

When the GET method is used, the parameter \texttt{source} must be provided
to indicate where semantic data source can be found. In the other hand, when POST method is used, the parameter \texttt{source} is not necessary because the semantic data must be contained by the HTTP message. If the parameter \texttt{provider} is not provided, VPOET will return the \textquotedblleft best visualisation\textquotedblright  given the user profile pointed by parameter \texttt{userProfile}.  When there is no template for a requested element, a default visualisation is provided.

An Fortunata-based application, called MIG,  provide users with a form (in a wiki page) to specify the user profile. As any Fortunata-based application, this information is public and accessible.

The HTTP messages with the specified syntax can be sent to VPOET by other programs (agents) written in any programming language, or by javascript applications executed in a web browser. However, browsers are more limited than other applications because they suffer security restrictions due to communication is restricted to the server which holds the web application. However, our approach do not have this problem because communications are centralised by Fortunata.

\begin{figure*}[t]
    \centering
    \includegraphics [width=4in]{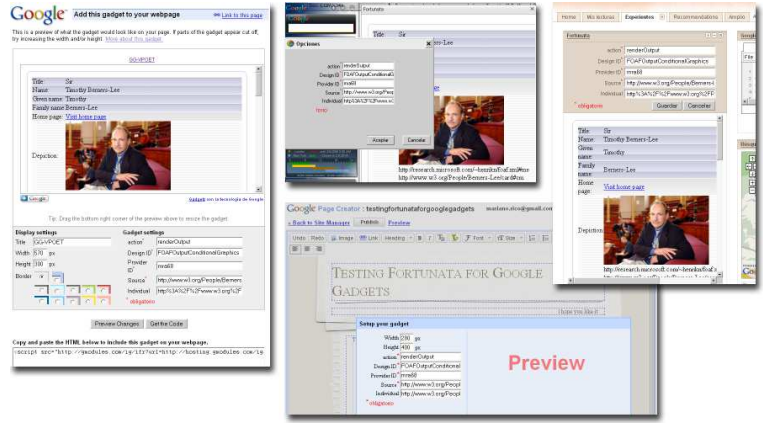}
    \caption{Using GG-VPOET in different application oriented to end-users. In clockwise:  a personal page, Google Desktop, iGoogle, and Google Pages}
    \label{fig:xample_demo_GG}
\end{figure*}

To let final users exploit this channel, a Google Gadget has been implemented, as was show in figure \ref{fig:fortunata_parties}. In this figure, user4 use this gadget in its web pages, or in some Google products, such as iGoogle, Google Pages, or Google Desktop. This gadget is configured providing the same information that was provided for the test phase. Figure \ref{fig:xample_demo_GG} shows this gadget in action using an output template for foaf:Person.

\section{Matching the user profile and the VPOET semantic templates}
\label{sec:matching}
\begin{figure*}[t]
    \centering
    \includegraphics [width=4.6in]{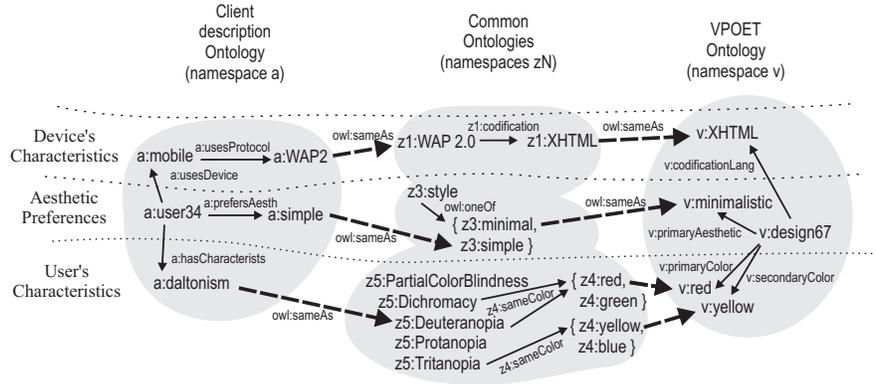}
    \caption{Matching process to find a visualisation template from a given user profile}
    \label{fig:semantic_promise}
\end{figure*}

Let us suppose that VPOET contains different templates for \texttt{foaf.Person}, and an external application requesting a \texttt{foaf.Person} template through the HTTP channel. VPOET should return \textquotedblleft the most adequate\textquotedblright  template for a given user profile. An example of this matching process is depicted in figure \ref{fig:semantic_promise}.

Each ontology, identified by a namespace, is shown as a cloud. The elements of the ontology, and their individuals, are shown inside its cloud; with ontology elements and some individuals inside the cloud. The left part of this figure shows the ontology describing the user profile, characterised by namespace $a$. In this example, the user identified as \textit{a:user34} has the following profile: (1) uses a WAP2 mobile phone as interaction device, (2) prefers simple aesthetics and (3) he/she is daltonic (colour-blindness associated to red-green colours).

In centre part of figure \ref{fig:semantic_promise}, public well-known ontologies are shown. Ontology $z_1$ indicates that the protocol WAP2.0 is codified as XHTML. For ontology $z_3$,\textquotedblleft minimal\textquotedblright  and \textquotedblleft simple\textquotedblright  are different kinds of styles but semantically close. Ontology $z_5$ has a visual-impairments hierarchy.

The right part of figure \ref{fig:semantic_promise} shows the VPOET ontology, with namespace $v$. In this ontology, the template identified as \textit{v:design67} is codified using the XHTML language, its primary aesthetic is minimalistic, and it has red and yellow as primary and secondary colours.

With just this semantic information, it is impossible to find that \textit{v:design67} is even a valid template for \textit{a:user34}. An additional semantic data source is required in order to link elements belonging to different ontologies. These links use to be \textquotedblleft sameAs\textquotedblright \footnote{Technically there are three types: owl:sameAs, owl:equivalentClass and
owl:equivalentProperty to distinguish individuals,
classes, and properties/relations respectively.}  relations, shown as discontinuous bold arrows in figure \ref{fig:semantic_promise}. Joining all this semantic information, a semantic agent can make a semantic query (e.g., using SPARQL language) based in the user profile, like this one: \textquotedblleft select a template with these characteristics: (1) codified in XTHML, (2) with minimalism as chief aesthetic, and (3) with primary colours avoiding red and green tones for text and background\textquotedblright . For this example, the result of this query would be the design \textit{v:design67}. Additional restrictions can refine the query to get the \textquotedblleft most adequate\textquotedblright  template for a given user profile.

\section{Conclusions and future work}
\label{sec:conclusions_futWork}
The work presented in this paper aims at providing developers with a simple and collaborative programming framework i order to simplify the process of creation of semantic web applications. Developers require (1) development environments simple and collaborative, (2) facilities for reuse of the contributed functionality, and (3) minimal dependencies between contributors. To achieve these requirements, we have taken advantage of an open source wiki-engine. We have developed a java library called Fortunata in order to facilitate developers the creation of plugins with semantic capabilities. As a proof-of-concept, some applications have been built using Fortunata. VPOET is an example of one of these applications.

From a developer's perspective, we consider that the targeted requirements concerning developers are successfully accomplished by the selected wiki-engine. However, it must be noticed that this is the result of our experience for some concrete applications. Concerning end-users, these applications are intended for a wide audience with no previous training in programming or semantic web technologies. This objective has been achieved be means of forms and simple macros, and experiments with end-users (not described in this paper) confirm it.  

These are the initial steps towards a semantic agent capable of providing an automatic generation of the user interface. This agent can use the data provided by VPOET in order to adapt the user interface to the user's profile (device used, user's impairments, and aesthetic preferences). Many open aspects remains open: composition of templates, or interaction between templates, among others. The architecture shown in this paper can provide developers with a simple but powerful infrastructure to achieve these long-term objectives.

%
%
%





\printindex

\end{document}